\documentclass[reprint,aps,prd,superscriptaddress,preprintnumbers,nofootinbib]{revtex4-1}

\usepackage{mathrsfs} 
\usepackage{amsfonts}
\usepackage{amsmath}
\usepackage{amsthm}
\usepackage{amssymb}
\usepackage{array}
\usepackage{dcolumn}
\usepackage[normalem]{ulem}
\usepackage[svgnames]{xcolor}
\usepackage[hidelinks]{hyperref}
\hypersetup{
    colorlinks=true,
    linkcolor=NavyBlue,
    filecolor=NavyBlue,
    urlcolor=NavyBlue,
    citecolor=NavyBlue,
    }

\usepackage{graphics}
\usepackage{graphicx}
\usepackage[caption=false]{subfig}
\usepackage{adjustbox}
\usepackage{breqn}
\usepackage{fontawesome} 
\usepackage{slashed}

\usepackage{tikz}
\usetikzlibrary{shapes.geometric, arrows}

\tikzstyle{startstop} = [rectangle, rounded corners, minimum width=3cm, minimum height=1cm,text centered, draw=black, fill=red!30]

\tikzstyle{process} = [trapezium, trapezium left angle=70, trapezium right angle=110, minimum width=3cm, minimum height=1cm, text centered,text width = 3cm, draw=black, fill=blue!30]

\tikzstyle{constraint} = [rectangle, minimum width=3cm, minimum height=1cm, text centered, text width = 3cm, draw=black, fill=orange!30]

\tikzstyle{decision} = [diamond, minimum width=3cm, minimum height=1cm, text centered,text width = 3cm, draw=black, fill=green!30]

\tikzstyle{arrow} = [thick,->,>=stealth]

\newcommand\lsim{\lesssim}

\def\be{\begin{equation}}
\def\ee{\end{equation}}
\def\bea{\begin{eqnarray}}
\def\eea{\end{eqnarray}}

\def\eff{\rm eff}

\newcount\Comments  
\newcommand{\kibitz}[2]{\ifnum\Comments=1\textcolor{#1}{#2}\fi}

\makeatletter
\let\cat@comma@active\@empty
\makeatother

\begin{document}

\hfill \preprint{LA-UR-26-21581}

\title{Self-Interaction and Galactic Magnetic Field Bounds on Millicharged Magnetic Monopole Dark Matter}

\author{Michael L. Graesser}
\affiliation{Theoretical Division, Los Alamos National Laboratory, Los Alamos, NM 87545, USA}

\author{R. Andrew Gustafson}
\affiliation{Theoretical Division, Los Alamos National Laboratory, Los Alamos, NM 87545, USA}
\affiliation{Center for Neutrino Physics, Department of Physics, Virginia Tech, Blacksburg, Virginia 24601, USA}

\begin{abstract}

A dark matter sector composed of magnetic monopoles of a dark U(1) symmetry having a small kinetic mixing with the Standard Model photon has a rich and interesting phenomenology. The model in itself is also of theoretical interest. Based on the temperature of the dark sector and scale of spontaneous symmetry breaking for this U(1), three phenomenologically distinct cases for this model of dark matter are discussed. In all cases,  constraints on dark matter self-interactions are translated into constraints on the model parameters.  As the magnetic monopoles acquire a small visible magnetic charge,  the survival of galactic magnetic fields, known as the Parker effect, places further constraints on the mixing between the dark and visible sectors.
\end{abstract}

\maketitle

\section{Introduction}

Magnetic monopoles occur naturally in gauge theories where a scalar vacuum expectation value (vev) spontaneously breaks an $SU(2)$ symmetry down to a $U(1)$ \cite{tHooft:1974kcl,Polyakov:1974ek}. If the resulting $U(1)$ is spontaneously broken by another scalar vev, then the monopoles are connected by flux strings to monopoles of opposite magnetic charge (hereafter referred to as ``antimonopoles") \cite{Nielsen:1973cs,Nielsen:1973qs,Nambu:1974zg}.

As the Standard Model (SM) photon is massless, we do not expect the scenario of confinement if magnetic monopoles with visible magnetic charges exist. However, it is possible that these monopoles exist in a dark sector \cite{Terning:2018lsv, Terning:2018udc, Terning:2019bhg, Terning:2020dzg}, with a dark $U(1)$ kinetically mixed with the SM photon \cite{Okun:1982xi, Holdom:1985ag}. In this scenario, the dark monopoles can now have some small effect on SM particles. 
As such, we might consider these ``millicharged magnetic monopoles", although the interplay between the dark and visible $U$(1)s is more nuanced than simply rescaling the magnetic charge.

Dark magnetic monopoles of a dark sector kinematically mixed with the SM photon has also attracted interested in itself as a toy
model for suggesting a plausible resolution of certain problems in the physics of magnetic monopoles. A long time ago Weinberg noticed that electron-monopole scattering is not only non-perturbative (because the tree-level amplitude is 
proportional to $eg \sim 4 \pi)$, but also that the scattering amplitude is not Lorentz-invariant \cite{Weinberg:1965nx}. 
In contrast, dark monopole - electron scattering is perturbative in the small mixing parameter $\epsilon$, and after summing over soft dark and SM photon emissions the otherwise Lorentz violation in the amplitude is exponentiated into a phase which becomes unity after applying Gauss' linking number formula and the Dirac quantization condition \cite{Terning:2018udc}.

We consider a situation where  millicharged monopoles comprise the entirety of dark matter. Monopoles and antimonopoles are required to have different flavors so that stable bound states can be formed, but the monopole masses can be the same regardless of flavor (this is the same scenario explored in \cite{Terning:2019bhg}). Only energies far below the monopole mass are considered, so we are agnostic as to its production and the exact symmetry breaking that creates it. The breaking of the dark U(1) is considered, as this behavior gives rise to 3 different phenomenological cases: \textbf{A}: the dark temperature restores the symmetry, \textbf{B}: the symmetry is broken and the monopoles form Coulomb-like states, and \textbf{C}: the symmetry is broken and the monopoles form string-like states.

This paper is organized as follows. In Sec. \ref{sec:Model} the model of millicharged magnetic monopoles is briefly described. Then our constraints of interest are described in Sec. \ref{sec:DM_Bounds}. Three distinct cases for this model and the corresponding bounds are considered in Sec. \ref{sec:Cases}. Complementary constraints are shown in Sec. \ref{sec:Comp_Const}. A discussion of direct detection, namely the difficulty of performing such searches in this model, is provided in Sec. \ref{sec:Difficulty-with-Direct-Detection}. Finally, we conclude in Sec. \ref{sec:Conclusion}.

\section{Model \label{sec:Model}}
A new $U(1)$ gauge symmetry is considered which has a corresponding boson $A^{\prime}$, which we will call the ``dark photon" with gauge coupling $e_{D}$. There are also monopoles $M$ in this sector which have magnetic coupling $g_{D} = 4 \pi/e_{D}$, satisfying Dirac charge quantization. The behavior of monopoles and charges can be equivalently described in a two-potential formalism \cite{Zwanziger:1970hk} or by a single potential with Dirac strings connected to the monopoles \cite{Dirac:1948um}. In either case, the corresponding Maxwell's Equations can be written as

\begin{equation}
    \begin{split}
    \partial_{\nu} F^{ \prime\mu \nu} &=  J_{D}^{\mu} \\
    \partial_{\nu} \Tilde{F^{\prime}}^{\mu \nu} &=  K_{D}^{\mu}
    \end{split}
\nonumber
\end{equation}
where $J_{D}^{\mu}$ and $K_{D}^{\mu}$ are the dark electric and magnetic currents respectively (with the coupling constants absorbed into their definition), and $F^{\prime \mu \nu}$ ($\Tilde{F^{\prime}}^{\mu \nu}$) is the field strength tensor (dual field strength tensor).

An additional complex scalar field $\phi$ which is charged under the new $U(1)$ is added with the Lagrangian
\begin{equation}
    \mathcal{L}_{\phi} = |D_{\mu} \phi|^2 + \mu^{2} |\phi|^2  - \frac{\lambda}{2} |\phi|^4,
   \nonumber 
\end{equation}
where $D_{\mu} = \partial_{\mu} + i e_{D} A^{\prime}_{\mu}$. The dark Higgs $\phi$ has a vacuum expectation value of $\langle \phi \rangle = \mu/\sqrt{\lambda}$. When the field takes this vev, the dark photon acquires a mass $m_{A^{\prime}} = \sqrt{2} e_{D} \langle\phi \rangle$. This non-zero vev also causes monopole-antimonopole pairs to be confined by Nielson-Olsen flux strings \cite{Nielsen:1973cs,Kobayashi:1974ph}. Thus, a non-relativistic Hamiltonian for these monopole-antimonopole bound states is given by \cite{Terning:2019bhg}

\begin{equation}
    H = \frac{p^2}{2 \mu_{M}} - \frac{\alpha_{g}}{r}\exp(-m_{A^{\prime}}r) + C \pi \langle \phi \rangle^2 r \label{eq-hamiltonian}
\end{equation}
where $C$ is some $\mathcal{O}(1)$ number, $r$ is the separation distance, and $\mu_{M}$ is the reduced mass of the pair ($\mu_{M} = m_{M}/2$ for the case of equal mass monopoles and antimonopoles). We have introduced $\alpha_{g} = g_{D}^2/4\pi$ as an analog for the electromagnetic fine-structure constant. For small values of $\langle \phi \rangle$ the state is essentially Coulomb-like, while for large $\langle \phi \rangle$, the state is dominated by the tension term and is described spatially by Airy-functions.

If the dark sector has a temperature $T_{D} \gtrsim \mu / \sqrt{\lambda}$, then $\langle \phi \rangle \rightarrow 0$. This temperature also induces a background of dark radiation comprised of $A^{\prime}$, $\phi$ and $\phi^{*}$. Thus, while in this circumstance the dark photon does not acquire a mass from the $\phi$ vev, it does obtain a temperature dependent effective mass $m_{A^{\prime} \eff} \sim \mathcal{O}(T_{D})$ number \cite{Berlin:2022hmt}. \footnote{The authors of \cite{Berlin:2022hmt} consider a weakly coupled dark radiation, which is different than the case here as we exclusively consider the dark sector magnetic coupling to be perturbative ($e_{D} > 4 \pi$). We do not re-derive the full equations of motion, and instead claim that the effective mass and temperature are connected at the $\mathcal{O}(1)$ level.}

The similarities of the Hamiltonian in Eq.~(\ref{eq-hamiltonian}) to that of hydrogen prompt us to consider ionization. However, the last term prevents the monopoles from ever truly being free so long as $\langle \phi \rangle >0$. Given enough energy, the excitation number becomes large and the separation between the monopoles can reach macroscopic lengths, making it effectively ``ionized". We touch on the nuances of this for the different scenarios in Sec. \ref{sec:Cases}.

The kinetic mixing interaction between the visible and dark sectors is given by 
\begin{equation}
    \mathcal{L} \supset \varepsilon F^{\mu \nu} F^{\prime}_{\mu \nu} \, ,
    \nonumber
\end{equation}
where $F^{\mu \nu}$ is the field strength tensor for the SM photon. Through a field redefinition $A_{\mu} \rightarrow A_{\mu} + \varepsilon A^{\prime}_{\mu}$, $A^{\prime}_{\mu} \rightarrow A^{\prime}_{\mu}$, the kinetic mixing of the fields can be undone, giving visible and dark Maxwell's equations in the form of \cite{Terning:2018lsv, Hook:2017vyc, Graesser:2021vkr}
\begin{equation}
    \begin{split}
        \partial_{\nu}F^{\mu \nu} &=  J^{\mu} \\
        \partial_{\nu} \Tilde{F}^{\mu \nu} &= - \varepsilon K^{\mu}_{D} \\
        \partial_{\nu} F^{\prime \mu \nu} &= m_{A^{\prime}}^{2} A^{\prime \mu} + J_{D}^{\mu} + \varepsilon J^{\mu} \\
        \partial_{\nu} \Tilde{F}^{\prime \mu \nu} &= K_{D}^{\mu}
    \end{split}
    \nonumber
\end{equation}
where $m_{A^{\prime}}$ can apply to the dark photon mass from symmetry breaking, or from the non-zero temperature. It is assumed that there are no magnetic monopoles in the visible sector.

Under these field redefinitions, the visible electric currents have obtained an $\varepsilon$ suppressed coupling to the dark photon, and the dark monopoles have obtained a sign-flipped and $\varepsilon$ suppressed coupling to the visible photon. The effective coupling between the SM electric charges and the dark monopole separated by length $r$ will scale as $\sim 1 - \exp(-m_{A^{\prime}}r)$. In terms of Feynman diagrams, the first term arises from the exchange of the SM photon, while the second from the exchange of the dark photon. Thus, at energies larger than $m_{A^{\prime}}$ (or lengths shorter than $(m_{A^{\prime}})^{-1}$), there is effectively no coupling between visible charges and dark monopoles, while at lower energies (longer lengths) the dark photon can no longer contribute and the monopoles have a non-vanishing magnetic charge in the visible sector \cite{Hook:2017vyc, Terning:2018lsv}.

\section{Dark Matter Bounds \label{sec:DM_Bounds}}

\subsection{Self-Interactions \label{sec-self-int}}

Observations of dark matter distributions place constraints on the strength of dark matter self-interactions. For 2-2 velocity independent scattering, galactic simulations indicate that dark matter sub-structure is consistent with $\sigma/(2m_{M}) \lesssim 2 \times 10^{-25} \mathrm{cm^{2}}/\mathrm{GeV}$ \cite{CosmicSIDM1, CosmicSIDM2}. We can utilize this constraint for bound state-bound state scattering, as the cross section is fixed by the geometric size of the state.

For ionized monopoles, the long-range interactions can induce plasma instabilities which lead to momentum transfer far larger that expected for 2-2 scattering \cite{Lasenby:2020rlf,Cruz:2022otv,DeRocco:2024ifs}. In the limiting case where the dark photon is massless, a limit of $(g_{D} m_{p})/(e m_{M}) \lesssim 2 \times 10^{-14}$ is set for a fully ionized dark sector \cite{Cruz:2022otv,DeRocco:2024ifs}, where $m_{p}$ is the mass of the proton. This is a strong constraint on the coupling, but as it is based on gravitational observations, it is only valid for ionization fractions $ f_{D} \gtrsim 0.1$. (Here $f_{D}$ is defined to be the ratio of \textit{free} positively charged monopoles to the \textit{total} (both bound and free) negatively charged monopoles.)

We note also that these previous analyses did not include the presence of dark radiation. As there is a non-perturbative coupling between $\phi$ and monopoles, the collective behavior of ionized monopoles might be significantly altered. Here, we do not re-derive plasma instabilities in this strongly coupled dark matter-dark radiation scenario in this work. Rather, we show in our exclusion where $f_{D} \gtrsim 0.1$ as a hatched region indicating an expected (but not certain) exclusion on those parameters.

\subsection{Galactic Magnetic Field Survival \label{sec:Parker}}

A large enough population of SM magnetic monopoles can wreak havoc on the magnetic fields of galaxies at large scales. The pioneering work of Parker shows that in the presence of a magnetic field, magnetic monopoles are accelerated and can drain the stored energy of the magnetic field on timescales shorter than the dynamo regeneration time, provided the monopole number density is high enough.
Known as the Parker effect, this process strongly limits the number density of monopoles \cite{Parker:1970xv,Turner:1982ag}. 

Dark magnetic monopoles can also dissipate SM magnetic fields, but the analysis is slightly more involved. Dark magnetic monopoles are confined when the dark photon is massive, and this affects how they couple to SM electric charges and currents. This leads to an additional distance dependence, as mentioned previously. Additionally, the ground state of bound dark monopole-antimonopoles has no magnetic properties, and therefore classically the ground state doesn't dissipate SM magnetic fields. 
In an external SM magnetic field however, 
quantum mechanically the ground state is unstable, provided the tension is small enough compared to magnetic field. After escaping the Coulomb barrier, the dark monopoles can then dissipate the magnetic field, leading to stringent constraints on $Q_{\rm eff}$ \cite{Graesser:2021vkr}. 

Compared to that work, here we consider how monopole-monopole scattering can effectively ionize dark magnetic monopoles, which then go on to dissipate galactic magnetic fields. To begin, a fraction of dark magnetic monopoles is assumed to be ionized, an assumption that is evaluated in Sec.~\ref{sec:Cases}.

Following an approach similar to \cite{Turner:1982ag}, Newton's Law for non-relativistic monopoles in a constant magnetic field $\mathbf{B}$ is
\begin{equation}
    \frac{d \mathbf{v}}{dt} = \frac{Q_{\eff}}{m_{M}} \mathbf{B} \label{eq-EOM},
\end{equation}
where $Q_{\eff} = g_{D} \varepsilon$. The change to the kinetic energy $K$ gained by a monopole is
\begin{equation}
    \frac{dK}{dt} = Q_{\eff} \mathbf{B} \cdot \mathbf{v}.
    \nonumber
\end{equation}
As the initial velocity distribution is taken to be isotropic and there is an equal number of monopoles and anti-monopoles, the average value of this energy change is zero. Taking the second derivative of $K$ and using the equation of motion given by Eq.~(\ref{eq-EOM}) gives 

\begin{equation}
    \frac{d^{2} K}{dt^2} = \frac{Q_{\eff}^2 B^{2}}{m_{M}}.
    \nonumber
\end{equation}

The total energy change for a monopole which spends time $\Delta t$ in the coherent region is $\frac{1}{2} (\Delta t)^2 d^2 K/dt^{2}$. The average time for these monopoles to pass through a coherent magnetic field of length $R_{coh}$ is $R_{coh}/v_{M}$, where $v_{M} \sim 10^{-3}$ is the typical monopole velocity. Therefore, if there is a flux of monopoles $\mathcal{F}$ (in units of $\mathrm{cm^{-2} s^{-1} sr^{-1}}$), then the amount of energy loss from a coherent region (assumed to be a sphere) is
\begin{equation}
    \frac{dE}{dt} = 8 \pi^2 \mathcal{F} R_{coh}^2 \times \frac{(\Delta t)^2}{2} \frac{d^2 K}{dt^2} = \mathcal{F}\frac{4 \pi^2 R_{coh}^4 Q_{\eff}^2 B^2 }{m v_{M}^2}
    \nonumber
\end{equation}

The time $\tau$ at which the total energy loss becomes comparable to the energy stored in the fields $\sim 4 \pi R_{coh}^3 B^2/3$ is then given by 
\begin{equation}
    \tau = \frac{m_{M} v_{M}^2}{3 \pi \mathcal{F} R_{coh} Q_{\eff}^2}
    \nonumber
\end{equation}
where
\begin{equation}
    \mathcal{F} \simeq \frac{\rho}{4 \pi m_{M}} v_{M} \times f_{D}.
    \nonumber
\end{equation}
For this time to be longer than the time of the Milky Way dynamo $\tau_{dyn} \sim 10^{8}$ years constrains $Q_{eff}$ to be less than 
\begin{equation}
    Q_{\eff} \lesssim \sqrt{\frac{4 m_{M}^2 v_{M}}{3 \rho_{DM} f_{D} R_{coh} \tau_{dyn}}},
    \nonumber
\end{equation}
which leads to a constraint on $\varepsilon$. 

Turning back to SM monopoles, one could object that the conclusions of the Parker effect are premature, and instead entertain the possibility that the galactic magnetic field is generated by monopole plasma oscillations \cite{PhysRevLett.50.544}. If so, the vanishing of the large-scale galactic field is but the first phase of a cycle of an oscillation in the magnetic field. Parker shows however that this scenario has significant challenges, as it leads to peculiar results at odds with observations: Alfen waves travel at $\big(\sqrt{2} \big)^{-1}$ of their textbook value; and the velocity of the magnetic field is $1/2$, instead of the same, as the ideal electrically conducting fluid \cite{1987ApJ...321..349P}.

\section{Phenomenological Cases \label{sec:Cases}}

The phenomenology of this model depends on the relative sizes of different parameters and energies. While we can expect couplings/masses to be effectively constant, the temperature will change over the cosmic history of interest. The present-day dark radiation temperature $T_{D}$ is related to that at different redshifts $z$ through
\begin{equation}
    T_{D}(z) \simeq (1+z) T_{D}
    \nonumber
\end{equation}
that is, assuming that over the time of interest there are no additional sources of entropy transfer in the thermal bath. We will often relate this temperature to the present-day temperature of the Cosmic Microwave Background, which we denote as $T_{\gamma}$.

\subsection{Symmetry Restoration}

In this case, the temperature of the dark sector, $T_D$, is assumed to satisfy
\begin{equation} 
T_{D} \gg  \mu /\sqrt{\lambda} ~. 
\nonumber 
\end{equation}
This drives $\langle \phi \rangle \rightarrow 0$ for all relevant times, so monopoles are unconfined. 
The ionization fractions will be shown below to depend on the relationship between $T_{D}$ and $\alpha_{g}^2 m_{M}$. 

Consider first the bound state, which is Coulomb-like since in this case the tension vanishes. The geometric cross section is
\begin{equation}
    \sigma_{\mathrm{geo}, C} = \frac{\pi}{\alpha_{g}^2 m_{M}^2},
    \nonumber
\end{equation}
which can be used to set an exclusion as discussed in Sec.
\ref{sec-self-int}. 

Due to the non-perturbative coupling between the $\phi$ and the monopoles, the dark matter is in thermal equilibrium with the dark radiation at all times. As such, the ionization fraction at early times can be approximated using the Saha Equation
\begin{equation}
    \begin{split}
        \frac{f_{D,\mathrm{Saha}}^2(z)}{1 - f_{D,\mathrm{Saha}}(z)} = \frac{2 m_{M}}{\rho_{DM}(z)} \bigg[& \bigg(\frac{m_{M} T_{D}(z)}{4 \pi}\bigg)^{3/2} \\ & \times \exp (- B_{D}/T_{D}(z) ) \bigg]
    \end{split}
    \nonumber
\end{equation}
where  $B_{D}  = \alpha_{g}^2 m_{M}/4$ is the binding energy for the monopole-antimonopole state. The dark matter density evolves as
\begin{equation}
        \rho_{DM}(z) = (1 + z)^3 \rho_{DM}(z = 0) ~.
        \nonumber
\end{equation}

Next consider the volumetric rate (the thermally averaged recombination cross-section) (see Eq 2.3 in the published version of \cite{Kaplan:2009de})
\begin{equation}
 \mathcal{A}(T_{D}(z)):= \langle \sigma v \rangle  = \xi \frac{64 \pi}{\sqrt{27 \pi}} \frac{\alpha_{g}^2}{\mu_{M}^2} x^{1/2} \log  x
    \nonumber
\end{equation}
where $x=B_D/T_D(z)$ and $ \xi = 0.448$ is a best fit parameter for massless dark photons \cite{Ma:1995ey} (as we assume the lowest-lying states are Coulomb-like, any effective-mass induced corrections are small). Note that this rate is only non-negative for $B_{D} > T_{D}$, meaning that the states will only start to recombine once the temperature drops below the binding energy.

Next an equilibrium redshift $z_{eq}$ is defined at which the Hubble rate $H(z)$ equals the recombination rate, that is
\begin{equation}
    f_{D, \mathrm{Saha}}(z_{eq}) \frac{\rho_{DM}(z_{eq})}{2 m_{M}} \mathcal{A}(z_{eq}) = H(z_{eq}). \label{eq-Equilibrium-z}
\end{equation}
After this time the ionization fraction remains fixed and provides an estimate of the present-day ionization fraction
\begin{equation}
    f_{D}(z = 0) = f_{D,\mathrm{Saha}}(z_{eq}).
    \nonumber
\end{equation}

\begin{figure}
    \includegraphics[width = 0.45 \textwidth]{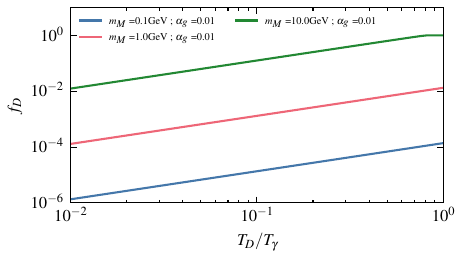}
    \caption{Case \textbf{A}. Present-day ionization fraction as a function of the present-day dark temperature-to-photon temperature ratio, in the case where the temperature drives $\langle \phi \rangle \rightarrow 0$. The behavior for 3 different monopole masses are displayed. \label{fig-ion-v-temp}}
\end{figure}

In our analysis if the Hubble rate is found to be always greater than the recombination rate then the ionization fraction is set to 1 (while this might miss some phenomenology due to a small number of bound states, we will see that such large ionization fractions are expected to be excluded). The dependence of the ionization fraction on the dark photon temperature is shown in Fig. \ref{fig-ion-v-temp}. \footnote{Although there is no closed form solution for $f_{D}$, we can estimate it as $f_{D} \simeq C_{D} (m_{M}/\alpha_{g}^2)^2 (T_{D}/T_{\gamma})$ for $T_{D} < B_{D}/30$ where $C_{D}\simeq 1.3 \times 10^{-10}$, and $f_{D} \simeq 1$ for $T_{D} > B_{D}/30$. } 

\begin{figure*}
    \includegraphics[width = 0.45 \textwidth]{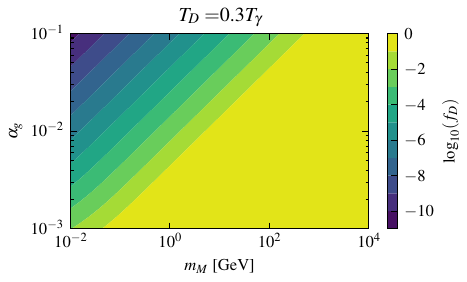}
    \includegraphics[width = 0.45 \textwidth]{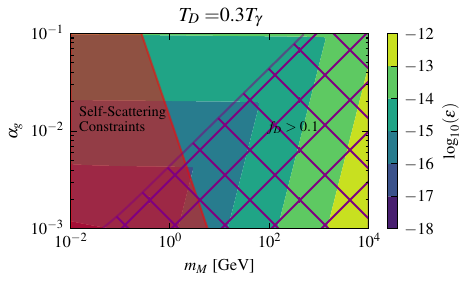}
    \caption{Case \textbf{A}. \textbf{Left:} Ionization fraction at present times for monopoles as a function of their mass and coupling for a dark temperature $T_{D} = 0.3 T_{\gamma}$ . \textbf{Right:} Constraints on $\varepsilon$ due to the survival of galactic magnetic fields. Also shown are self-scattering constraints from bound-state 2-2 scattering, and the region where $f_{D} > 0.1$. In this latter case, we leave it hatched as it is expected that dark plasma effects \cite{Lasenby:2020rlf, Cruz:2022otv, DeRocco:2024ifs} and interactions with the dark radiation will be sizable, but the exact behavior is currently unknown. 
    Stellar cooling bounds \cite{An:2013yfc, An:2013yua} constrain 
    $\varepsilon \lesssim 8 \times 10^{-15}/\min(e_{D}, 4 \pi)$ provided $m_{A'} \lsim $ keV, as discussed in Sec. \ref{sec:Comp_Const}.
    \label{fig-Dark-CMB} }
\end{figure*}

Using the ionized fraction as input, the galactic magnetic field bound is then evaluated, as discussed in Sec. \ref{sec:Parker}.  Fig. \ref{fig-Dark-CMB} shows the present-day ionization fraction, along with exclusions. The ``hatched" region indicates the part of parameter space with $f_{D} \gtrsim 0.1$, for which reasonably sizable self-interactions is expected, as discussed previously in Sec. \ref{sec-self-int}.

\subsection{Coulomb dominated limit}
In this case 
the temperature for the dark sector is assumed to be low, that is
\begin{equation}
    \mu / \sqrt{\lambda} \gg T_{D}.
    \nonumber
\end{equation}
Under this condition, $\phi$ obtains an non-zero vev of $\langle \phi \rangle = \mu/\sqrt{\lambda}$ causing the monopoles to be confined. Because of the assumption on the temperature, the dark radiation is insufficient to cause any substantial ionization. However, self-interactions within galaxies may lead to a non-zero ionized fraction.\footnote{In principle, scattering could also lead to a number of excited states, but assuming that the lowest-lying states are Coulomb-like, $E_{n} \simeq E_{0}/n^2$ (where $E_0$ is the energy of the ground state), so ionization only requires marginally more energy than excitation.}

Next, the parameters of the potential are assumed to satisfy
\begin{equation}
 \alpha_{g}^{3/2} \mu_{M} \gg  \sqrt{C \pi} \mu /\sqrt{\lambda}
    \label{eqn:Coulomb-limit}
\end{equation}
such that the ground and first few excited states are atomic-like. This inequality expresses the assumption that on the scale of a Bohr radius $a_0 = 1/\alpha_g \mu_M$, a Coulomb potential dominates over the string tension potential. One can confirm that the condition for the Yukawa potential in Eq.~(\ref{eq-hamiltonian}) to be well-approximated by a Coulomb potential on distances of order the Bohr radius and smaller, namely $m_{A'} \ll \alpha_g  \mu_M$, is equivalent to Eq.~(\ref{eqn:Coulomb-limit}). Thus only one and not two conditions need to be assumed in order for the ground and first few excited state to be atomic-like.   

For Coulomb-like states with relative kinetic energy greater than their binding energy, self-scattering can lead to states having a positive energy as defined by Eq.~(\ref{eq-hamiltonian}), which we refer to as ``ionization"\footnote{As noted earlier, because of the flux tube, the state is not truly ionized, but rather excited to a high energy level such that the monopole and antimonopole have a separation much larger than $m^{-1}_{A'}$.}. To estimate this rate, consider the product $N_{M} \sigma_{I}$, which is the ionization cross section times the number of free positively charged monopoles expected in the final state that arise from the interaction.  The analytical calculation for hydrogen in \cite{sheldon1965cros} can be closely approximated by a linear function of kinetic energy (see their figure 3),  giving a cross section as a function of relative velocity $v$ between the two bound states, 
\begin{equation}
    N_{M} \sigma_{I}(v) = \begin{cases}
        0 \, \, \mathrm{for} \, \, v/\alpha_{g} < \sqrt{1/2} \\
        \sigma_{\mathrm{geo},C} \bigg(\frac{v^2}{\alpha_{g}^2} - \frac{1}{2} \bigg) \, \, \mathrm{for} \, \, \sqrt{1/2} < v/\alpha_{g} < \sqrt{5/2} \\
        2 \sigma_{\mathrm{geo}} \, \, \mathrm{for} \, \, v/\alpha_{g} > \sqrt{5/2},
    \end{cases}
    \nonumber
\end{equation}
That is, the ionization cross-section rises linearly with excess kinetic energy ($m_{M} v^2/2 - \mu_{M} \alpha_{g}^2/2$), until it saturates to twice the geometric cross section, implying both states are ionized.

We can now compute the scattering-induced rate of change to the ionization fraction.

\begin{equation}
    \begin{split}
    \frac{d f_{D}}{dt}\bigg|_{scat} = \frac{(1 - f_{D})^2 \rho_{DM}}{2 m_{M}}  \int& d^{3} \mathbf{v_{1}} d^{3} \mathbf{v_{2}} f_{\chi}(\mathbf{v_{1}}) f_{\chi} (\mathbf{v_2}) \\
    &N_{M}\sigma_{I}(|\mathbf{v_{1}} - \mathbf{v_{2}}|) |\mathbf{v_{1}} - \mathbf{v_{2}|}
    \end{split}
    \nonumber
\end{equation}
where $f_{\chi}(\mathbf{v})$ is the velocity distribution of the dark matter. The velocity distribution is taken to be a truncated Maxwell-Boltzmann distribution
\begin{equation}
    f_{\chi}(\mathbf{v}) = A \exp \bigg( \frac{-|\mathbf{v}|^2}{v_{char}^2}\bigg) \Theta \big(v_{esc} -|\mathbf{v}| \big)
    \label{eq-vel-distribution}
\end{equation}
where $\Theta(x)$ is the Heaviside theta function and
\begin{equation}
    \begin{split}
    A^{-1} = \pi v_{char}^2 \bigg[& \sqrt{\pi}v_{char} \mathrm{Erf} \bigg(\frac{v_{esc}}{v_{char}} \bigg) \\ &-2 v_{esc} \exp \bigg(\frac{-v_{esc}^2}{v_{char}^2} \bigg) \bigg]
    \end{split}
    \nonumber
\end{equation}

Scattering may also induce recombination. However, making the assumption that the recombination cross section is also $\mathcal{O}(\sigma_{\mathrm{geo},C})$, this process will not become efficient until the ionization fraction becomes $\mathcal{O}(1)$, which is already excluded by constraints on self-interactions.

The ``ionized" states described here are not truly free ions, but rather are still connected with a string, similar to the idea of ``quirks" \cite{Kang:2008ea, Harnik:2011mv}. One might worry that these stringy states could de-excite back into closely bound states.

First consider the string breaking by forming a monopole-antimonopole pair. Since however the monopoles are assumed to be much more massive than the characteristic string energy ($\sim\langle \phi \rangle$), pair production of monopoles from the energy stored in the string is exponentially suppressed by $m_{M}^2/\langle \phi \rangle^2$ \cite{Kang:2008ea}.

Next, energy loss can occur through the emission of $\phi$ particles. We will follow the ``spin-down" reasoning from \cite{Harnik:2011mv}. By equating the kinetic and potential energy of the string, the characteristic length of the string in the absence of other ionized states is
\begin{equation}
    L_{sing} \sim m_{M} v^{2} / \langle \phi \rangle^2 \sim \mathrm{pc} \bigg(\frac{m_{M}}{\mathrm{GeV}}\bigg) \bigg( \frac{v}{10^{-3}}\bigg)^2 \bigg(\frac{10^{-10} \mathrm{eV}}{\langle \phi \rangle} \bigg)^2
    \label{eq-Lsing}
\end{equation}
With each oscillation the string emits $\mathcal{O}(\langle \phi \rangle)$ of energy. Therefore, the characteristic time for the energy to substantially change is given by
\begin{equation}
    \tau_{dec,sing} \sim \frac{m_M^2 v^3}{\langle \phi \rangle^{3}} \sim10^{24} \mathrm{s} \bigg(\frac{m_{M}}{\mathrm{GeV}}\bigg)^2 \bigg( \frac{v}{10^{-3}}\bigg)^3 \bigg(\frac{10^{-10} \mathrm{eV}}{\langle \phi \rangle} \bigg)^3.
    \label{eq-String-State-Dec-Time}
\end{equation}

The phenomenology changes when there is a large number of ionized states. First, in an analogue to cosmic strings, if two flux strings pass through each other, it is energetically favorable for the strings to intercommute, i.e. the monopoles swap partners. For cosmic-strings, it has been determined that the probability of intercommuting is nearly unity \cite{shellard1987cosmic}. Therefore, the length of the string is expected to go as
\begin{equation}
    L \sim \min \bigg\{ \frac{m_{M} v^2}{\langle \phi \rangle^2} , n_{M}^{-1/3} \bigg \}.
    \nonumber
\end{equation}

The spin down constraints can now be more accurately determined as

\begin{equation}
    \tau_{SD} = \frac{m_{M} v L}{\langle \phi \rangle }.
    \nonumber
\end{equation}

For a single monopole, the rate of intercommuting is $\sim n_{M} L^2 v_{\chi}$. If the length of the string is much larger than the radius, that is $L \langle \phi \rangle \gg 1$, then during the intercommutation a loop of length $\mathcal{O}(L)$ can readily form, and thus with $\mathcal{O}(L \langle \phi \rangle^2)$ in energy. From this a time frame $\tau_{loop}$ is defined wherein the energy loss due to the formation of loops is comparable to the kinetic energy of the monopole, namely
\begin{equation}
    \tau_{loop} \simeq \frac{m_{M} v_{\chi}^2}{(n_{M} L^2 v_{\chi})(L \langle \phi \rangle^2)} \simeq \frac{m_{M} v_{\chi}}{n_{M} \langle \phi \rangle^2} \max \bigg\{ n_{M} , \frac{\langle \phi \rangle^6}{m_{M}^3 v_{\chi}^{6}} \bigg \}
    \nonumber
\end{equation}

Finally, a total decay time $\tau_{dec}$ is defined as

\begin{equation}
    \tau_{dec} = \bigg(\frac{1}{\tau_{SD}} + \frac{\Theta(L \langle \phi  \rangle -1)}{\tau_{loop}} \bigg)^{-1} \label{eq-tau-dec-full}
    \nonumber
\end{equation}

Putting this all together gives a differential equation for the ionization fraction,
\begin{equation}
    \frac{d f_{D}}{dt} = \frac{d f_{D}}{dt}\bigg|_{scat} - \frac{f_{D}}{\tau_{dec}}
    \label{eq-Ion-diff-Eq}
\end{equation}

\begin{figure}
    \includegraphics[width = 0.45 \textwidth]{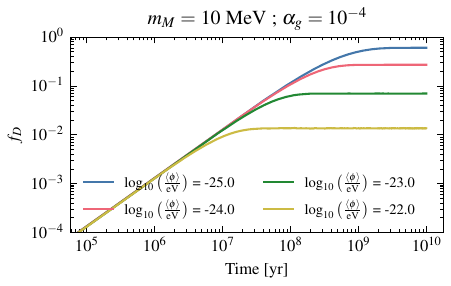}
    \includegraphics[width = 0.45 \textwidth]{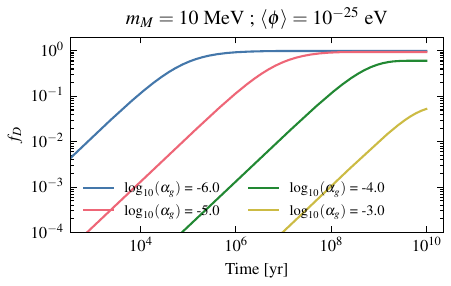}
    \includegraphics[width = 0.45 \textwidth]{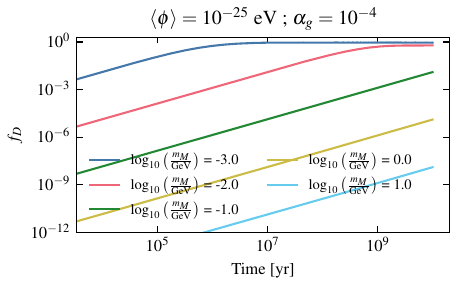}

    \caption{Ionization fraction as a function of time due to scattering (Case \textbf{B}), obtained by solving Eq.~(\ref{eq-Ion-diff-Eq}). In all calculations,  the dark matter density is fixed to $\rho_{M} = 0.3 \mathrm{\,GeV \, cm^{-3}}$, the escape velocity to $v_{esc} = 600$ km/s, and the characteristic velocity to $v_{char} = 220$ km/s as used in Eq.~(\ref{eq-vel-distribution}). The three different panels show the effects of varying $\langle \phi \rangle$ (\textbf{Top}), $\alpha_{g}$ (\textbf{Middle}), and $m_{M}$ (\textbf{Bottom}). The solutions are integrated up to $10^{10}$ years, a typical galactic lifetime. \label{fig-Ion-vs-Time}}
\end{figure}

\begin{figure}
    \includegraphics[width = 0.45 \textwidth]{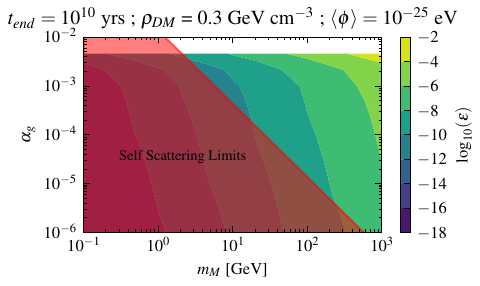}
    \caption{Case \textbf{B}. Values of $\varepsilon$ excluded from galactic magnetic field constraints on monopoles ionized through scattering . Also shown in red is parameter space excluded by bound-state self-interaction bounds. Stellar cooling bounds \cite{An:2013yfc, An:2013yua} constrain 
    $\varepsilon \lesssim 8 \times 10^{-15}/\min(e_{D}, 4 \pi)$ provided $m_{A'} \lsim $ keV, as discussed in Sec. \ref{sec:Comp_Const}. \label{fig-epsilon-bounds}}
\end{figure}

The evolution of this ionization fraction for several choices of parameters are shown in Fig. \ref{fig-Ion-vs-Time}. Setting the elapsed time to $10^{10}$ years, roughly the age of the Milky Way, then the ionized fraction can be used to obtain bounds on $\alpha_g$ and $\varepsilon$ the from the self-scattering and magnetic field observations discussed in Sec.~\ref{sec:DM_Bounds}, which are shown in Fig. \ref{fig-epsilon-bounds}.

\subsection{Tension dominated limit}
In this final case, the parameters satisfy the inequality
\begin{equation}
    \mu/\sqrt{\lambda} \gg \alpha_{g}^{3/2} m_{M},
    \nonumber 
\end{equation}
such that at a typical Bohr radius scale the Coulomb interaction is irrelevant compared to the string tension.
As with the previous case, the temperature is once again assumed to be small,
\begin{equation}
    \mu /\sqrt{\lambda} \gg T_{D} ~.
    \nonumber 
\end{equation}
The monopoles are confined and in a string-like state. The characteristic length and energy of the ground state are given by
\begin{equation} 
L_{g} \simeq (C\pi m_{M} \langle \phi \rangle^2)^{-1/3}, \quad E_g \simeq \frac{1}{m_M L^2_g}, 
\nonumber 
\end{equation} 
obtained from using dimensional analysis and the observation that $m_MH$ only depends on the single scale $(C \pi m_M \langle \phi \rangle^2)^{2/3}$ \cite{Terning:2019bhg}. 
Both the energy and the size of the n-th excited state scales as $n^{2/3}$.
The geometric size of the ground state is then 
\begin{equation}
    \sigma_{\mathrm{geo},S} \simeq \frac{1}{4} \bigg( \frac{\pi}{m_{M}^2 \langle \phi \rangle^{4}}\bigg)^{1/3}.
    \nonumber
\end{equation}
 As seen in the previous scenario, when the kinetic energy far exceeds the typical excitation energy, these states can be highly excited and reach macroscopic sizes. This excited state is closely described as a cylinder with length $L_{sing}$, as in Eq.~(\ref{eq-Lsing}), and radius $\sim \langle \phi \rangle^{-1}$ \cite{banks2008modern}. Unlike the Coulomb-like case, where interactions are primarily due to the exchange of dark photons by the monopoles, here the entire macroscopic string can interact with other states. As such, the cross section for an excited state to interact with and excite another ground state will go as the average projected area of the excited state

\begin{equation}
    \sigma_{g-ex} \simeq \langle \phi \rangle^{-1} L_{sing} \simeq \frac{m_{M} v_{M}^2}{\langle \phi \rangle^{3}}.
    \nonumber
\end{equation}

\begin{figure}
    \includegraphics[width = 0.45 \textwidth]{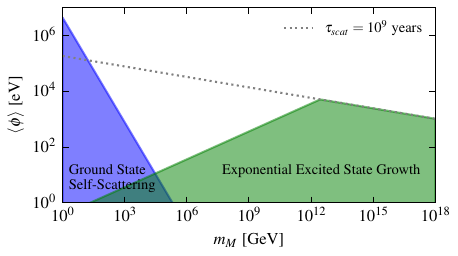}
    \caption{Case \textbf{C}. Bounds on string-like monopole states. Limits from self-interactions in the ground state are shown in blue, and in green is shown the parameter space where scattering should lead to an exponential increase in the number of excited states. This region also includes a requirement that scattering occurs on a time scale $\lesssim 10^{9}$ years, so there is significant time for an exponential increase to occur.  (The parameter space relating to this scattering time is shown as a dotted line.) A dark matter density of $\rho_{M} = 0.3 \mathrm{ \,GeV\, cm^{-3}}$ is assumed.  \label{fig-self-scat-string}}
\end{figure}

If the time between scatterings is less than the characteristic single-state decay time as given in 
Eq.~(\ref{eq-String-State-Dec-Time}), that is

\begin{equation}
    \frac{1}{\tau_{scat}}=\frac{1}{n_{M} \sigma_{g-ex} v_{M}} \lesssim \frac{m_{M}^2 v_{M}^2}{\langle\phi \rangle^3}, 
    \nonumber
\end{equation}
then each excited state will excite on average multiple other states, leading to an exponential growth in the excited fraction. Since these excited states have such a large cross section, a runaway increase in the number of excited states would violate self-interaction bounds. Thus, if the above inequality is satisfied and the typical scattering time is shorter than galactic timescales,  then this region leading to runaway growth is excluded.   Both the ground state self-scattering and exponential excited state growth are shown in Fig. \ref{fig-self-scat-string}.

While these monopoles do have a visible magnetic charge, there is no bound from the galactic magnetic fields. This is because the monopole-antimonopole pairs have a tension of $\sim \langle \phi \rangle$ between them. The magnetic force experienced by each of the monopoles is $\epsilon g_{D} B$, which must be larger than the tension for the Parker effect to occur. In other words, 
\begin{equation}
   \epsilon g_D B \simeq \epsilon g_D 10^{-6} \mathrm{G} \simeq \epsilon g_D 10^{-8} \mathrm{eV^2} \gtrsim \langle \phi \rangle ^2.
    \nonumber
\end{equation}
However, as can be seen by inspection from  Fig. \ref{fig-self-scat-string}, these values for the vev are already excluded.

\section{Complementary Constraints \label{sec:Comp_Const}}
This section provides an overview of previous constraints on this model. These constraints may involve all of the new particles introduced here, or only a subset.

The influence of millicharged magnetic monopole dark matter on galactic magnetic fields was previously explored in \cite{Graesser:2021vkr}. The authors of this previous work only considered very weakly bound monopole-antimonopole states with very small tension such that the magnetic field makes the state unstable and induces ionization through quantum tunneling. Our methods here of scattering- and dark radiation-induced ionization allows us to consider monopoles with much larger masses and coupling constants, as well as the situation where the string tension dominates over the Coulomb/Yukawa interaction.

In sufficiently strong magnetic fields, millicharged magnetic monopoles may be copiously produced via the Schwinger mechanism. This effect can be important for magnetars, which are neutron stars with magnetic fields $\mathcal{O}(10^{15} G)$. Insisting that the magnetic fields survive for $\gtrsim 10^{4}$ years sets strong constraints on models of light (sub-keV) monopoles with $m_{A^{\prime}}$ greater than the inverse length of the magnetar (roughly $(20 \mathrm{km})^{-1}$) \cite{Hook:2017vyc}. This bound does not require the monopoles to comprise dark matter. 

In principle, the magnetic field energy stored in magnetars could also be dissipated by a flux of relic dark magnetic monopoles. As is shown in Appendix \ref{app:DM-interactions-with-magnetars} however, the resulting bounds are weaker than the aforementioned Schwinger bounds or the self-interacting bounds, even under the conservative assumption that all the dark magnetic monopoles are ionized, and comprise all of the dark matter.

Many existing constraints exist on $A^{\prime}$ and $\phi$, even in the absence of magnetic-monopole dark matter. A recent review of dark photon limits can be found in \cite{Caputo:2021eaa} along with a regularly updated repository \cite{DPGithub}. Many of these bounds are agnostic as to the mechanism for dark photon mass, so they are independent of $\phi$ parameters.

Strong bounds on $\phi$ are obtained from stellar emission of a dark sector Higgs boson and dark photon through virtual exchange of a dark photon, causing cooling of stellar objects \cite{An:2013yfc, An:2013yua}. These authors found that for $\phi$ with masses below 1 keV, the ``effective coupling" $(\varepsilon e_D)$ of $\phi$ must satisfy $\varepsilon e_D \leq 8 \times 10^{-15}$. Naively, this coupling is just the product of $\varepsilon$ and $e_{D}$, although we must be more careful when $e_{D}$ is non-perturbative, which is the case here. For this work we consider $\alpha_g \in (10^{-6}, 10^{-1})$ which, because of the Dirac charge quantization condition, implies $e_D =\sqrt{4 \pi/ \alpha_g} \in ( \sim 10, \sim 3 \cdot 10^{3})$. The analysis of \cite{An:2013yfc, An:2013yua} is based on a tree-level calculation which for these $e_D$ couplings would have to be re-evaluated. The decay rate of the virtual dark photon into dark states is $\Gamma(m_{A'},T)=f(e_D,T/m_{A'}) m_{A'}/4\pi$ where, in this case at large $e_D$, $f$ is an unknown function that has to be evaluated non-perturbatively in $e_D$. At large couplings, if the dark photon is a broad resonance it is possible that in vacuum $f$ saturates at $4 \pi$, for then $\Gamma \sim m_{A'}$. Once the dark states are produced, 
the dark Higgs will have a prompt decay to a pair of dark photons (if kinematically allowed), while the dark photon still escapes the stellar object since it is long-lived -- it has an $\varepsilon$ suppressed decay to three visible photons.  Based on these observations we recast this limit as

\begin{equation}
    \varepsilon \lesssim \frac{8 \times 10^{-15}}{\min(e_{D}, 4 \pi)},
    \label{eqn:epsilon-min}
\end{equation}
while acknowledging that because of the non-perturbative nature of the dark Higgs sector, the actual limit could be somewhat different.

The existence of dark radiation, even without coupling to SM particles, will influence cosmic observables like Big Bang Nucleosynthesis (BBN) and the Cosmic Microwave Background (CMB). Bounds on the temperature can be related to bounds on the number of neutrinos $N_{\eff}$ 
\begin{equation}
    n_{D} \bigg(\frac{T_{D}}{T_{\gamma}} \bigg)^{4} = \frac{7}{4} \bigg(\frac{4}{11} \bigg)^{4/3}\Delta N_{\eff}
    \nonumber
\end{equation}
where $T_{\gamma}$ is the present-day temperature of the CMB and \sout{$g_{D}$} $n_{D}$ is the number of relativistic degrees of freedom in the dark sector (here we have  $n_{D} = 4$, 2 for $A^{\prime}$ and 2 for $\phi$). The reader is referred to Appendix \ref{sec:BBN} which provides more details of this comparison. Recent analyses of cosmic observables restrict $N_{\eff} \leq 0.124$ and $95\%$ confidence-level \cite{Fields:2019pfx}, implying $T_{D}/T_{\gamma} \leq 0.344$.

\section{Difficulty with Direct Detection}
\label{sec:Difficulty-with-Direct-Detection}

Thus far, we have not considered potential exclusions arising from monopole scattering in direct-detection experiments. These signals are suppressed in each of the three phenomenological scenarios, as we now describe.

\subsection{Symmetry Restoration}
If $\langle \phi \rangle \rightarrow 0$ due to the non-zero temperature, then the dark photon effective mass $m_{A^{\prime}, \eff} \simeq T_{D}$. Scattering with a momentum transfer $q$ greater than $m_{A^{\prime},\eff}$ is suppressed due to the opposite contributions of the dark and visible photons. Thus, the maximal energy deposited in the detector due to non-relativistic monopole-SM scattering will go as $T_{D}^2/m_{SM}$, much lower than the thresholds for most detectors.

\subsection{Coulomb-Dominated Limit}
In this case the condition $ \alpha_{g}^{3/2} m_{M} \gg  \mu /\sqrt{\lambda}$ implies that if $q < m_{A^{\prime}}$, then $q^{-1}$ is much larger than the size of the ground state. If so, the scattering does not resolve the bound state, and in the amplitude a cancellation occurs because of the opposite signs of the monopole and anti-monopole. Scattering between the ground state and Standard Model particles will be highly suppressed.

\begin{figure}
    \includegraphics[width = 0.45 \textwidth]{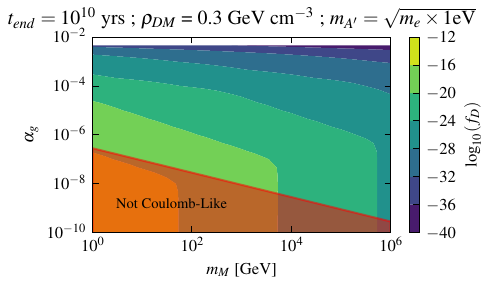}
    \caption{Scattering-induced ionization fraction as a function of $m_{M}$ and $\alpha_{g}$, taking a dark matter density of $0.3 \mathrm{GeV cm^{-3}}$ and a time evolution of $10^{10}$ years. We fix the dark photon mass to be $\sqrt{m_{e} \times 1 \mathrm{eV}}$, which implies that monopole-electron scattering can deposit $\sim$ 1 eV in the detector before suppression becomes large. In red is the region where the ground state is no longer Coulomb-like and is instead string-like. \label{fig-Ion-Frac-DD}}
\end{figure}

If the state is ionized, then the string length between the monopole and antimonopole can become much larger than $q^{-1}$ so the scattering can be treated as free. To have scattering which deposits at least the threshold energy $E_{th}$ in the detector requires
\begin{equation}
    \sqrt{m_{SM} E_{th}} \lesssim q \lesssim m_{A^{\prime}}.
    \nonumber
\end{equation}
where the inequality on the right is needed to avoid decoupling in the ultra-violet. For illustration purposes, consider setting the left and right side equal: $m_{A^{\prime}} = \sqrt{m_{SM} E_{th}}$ is fixed and the corresponding scattering-induced ionization fraction is obtained for $m_{SM} = m_{e}$ and $E_{th} =$1 eV, shown in Fig.~\ref{fig-Ion-Frac-DD}. 
By inspection, $f_{D} \lesssim 10^{-16}$, so published limits on dark matter cross sections would be 16 orders of magnitude weaker in this sitation. Thus, direct detection will be unable to probe small values of $\varepsilon$ in this scenario.

\subsection{Tension Dominated Limit}

\begin{figure}
    \includegraphics[width = 0.45 \textwidth]{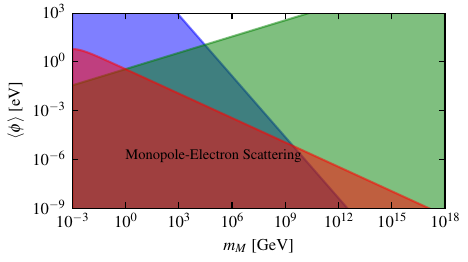}
    \caption{Parameter space for which monopole-electron scattering is possible for monopoles in string-like ground states. Also shown in blue and green are the self-interaction bounds (with the same color scheme as Fig. \ref{fig-self-scat-string}). \label{fig-String-DD}}
\end{figure}

For the ground-state, in order to resolve the monopole-antimonopole separation, it is required that $q \gtrsim L^{-1}_g \simeq (m_{M} \langle \phi \rangle^2)^{1/3}$, the inverse size of the ground state in the string dominated limit. From the kinematics of non-relativistic scattering, $q \lesssim \mu_{M-SM} v_{M}$ where $\mu_{M-SM}$ is the reduced mass of the monopole-Standard Model system. Therefore, to achieve a signal 
\begin{equation}
    \mu_{M-SM} v_{M} \gtrsim \big(m_{M} \langle \phi \rangle^{2} \big)^{-1/3} ~.
    \label{eq:ground_state_dd}
\end{equation}
However, any parameter space that satisfies this necessary inequality is already excluded due to self-interaction bounds (see Fig. \ref{fig-String-DD}). Ground state self-scattering alone rules out possible monopole masses up to $\sim 10^{9}$ GeV, and higher masses would imply exponential growth in the number of excited states. 

One could in principle consider highly excited states scattering in detectors, which, due to the macroscopic separation of the monopole and antimonopole, are not subject to the inequality in Eq.~(\ref{eq:ground_state_dd}). However, for these states, it is assumed that they either decay on fast time scales or else induce exponential growth of excitation for the monopoles, which is excluded through self-interaction bounds.

The case of string-like states with low excitation number remains phenomenologically interesting, especially in the case where the excitation energy is less than $\langle \phi \rangle$. In this case, decays via the emission of dark particles is energetically forbidden, making the excited state long-lived. This possibility has previously been explored by searching for the magnetic moments of these excited states \cite{Terning:2019bhg}. However, accurately calculating the lifetime of these states and subsequent scattering in detectors involves the interplay of all model parameters, and requires more involved calculations than those performed here. We will revisit this case in a future work.

\section{Conclusion \label{sec:Conclusion}}
This work considers a scenario in which dark matter is comprised of dark sector magnetic monopoles of a novel $U(1)^{\prime}$ gauge symmetry that is kinetically mixed with the Standard Model photon. There are 3 phenomenologically distinct possibilities: the monopoles are unconfined due to presence of dark radiation; are confined and Coulomb-like; or are confined and string-like. All of these cases are constrained here through self-interactions, and the first two can further be constrained by the survival of galactic magnetic fields.

Looking forward, this model can lead to interesting collider and cosmological effects. While we only looked at the structure impacts on galactic scales, the long-range interactions can lead to interesting structure formation, a direction explored in other models of long-range interactions \cite{Cyr-Racine:2012tfp,Giffin:2025oqe}. The dark sector magnetic monopoles could also have a potentially interesting impact on the cosmic microwave background radiation. It would also be interesting to explore cosmological origins for the dark sector magnetic monopoles, such topological dark matter \cite{Murayama:2009nj,Graesser:2020hiv}. Moreover, this model introduces symmetry breaking at different scales ($m_{M}$ and $\mu$), which might copiously produce strings and loops (i.e. closed strings). We plan to explore this further in a future work.

\section*{Acknowledgments}
The
work of MG is supported by the U.S. Department of Energy, Office of Science, Office of High Energy physics,
under contract number KA2401012 (LANLE83G) at Los
Alamos National Laboratory operated by Triad National
Security, LLC, for the National Nuclear Security Administration of the U.S. Department of Energy (Contract No.
89233218CNA000001).
This material is based upon work additionally supported by the U.S. Department of Energy, Office of Science, Office of Work force Development for Teachers and Scientists, Office of Science Graduate Student Research (SCGSR) program. The SCGSR program is administered by the Oak Ridge Institute for Science and Education (ORISE) for the DOE. ORISE is managed by ORAU under contract number DESC0014664. R.A.G. is a recipient of the SCGSR
Award. R.A.G is additionally grateful to the Virginia Tech Physics Department for continued support during this work.

\appendix

\section{BBN evaluation \label{sec:BBN}}
Consider the energy density in radiation $\rho_{R}$, which is composed of photons, electrons, neutrinos, and dark particles. These first three are in thermal equilibrium, so we have

\begin{equation}
    \frac{\rho_{R}}{\rho_{\gamma}} = 1 + \frac{7}{8} \bigg( \frac{n_{e}}{n_{\gamma}} + \frac{n_{\nu}}{n_{\gamma}} \bigg) + \frac{\rho_{D}}{\rho_{\gamma}},
    \nonumber
\end{equation}
where the factor of $7/8$ is because of the difference in fermion and boson statistics, and $n_{X}$ is the number of relativistic degrees of freedom for species $X$ ($n_{\gamma} = 2$, $n_{e} = 4$, $n_{\nu} = 2 N_{\nu}$). We can write $N_{\nu} = 3 + \Delta N_{\eff}$, where the Standard Model predicts $N_{\nu} = 3.044$ (this is different from 3 because of heating corrections \cite{Cielo:2023bqp}).

\begin{equation}
    \frac{\rho_{R}}{\rho_{\gamma}} = \frac{43}{8} + \frac{7}{8} \Delta N_{\eff} + \frac{n_{D} T_{D,BBN}^4}{n_{\gamma}T_{\gamma,BBN}^4},
    \nonumber
\end{equation}
where the $BBN$ subscript on the temperature means it applies to the time of BBN. Temperatures without additional subscripts are assumed to be present-day.

Typical bounds on the radiation during BBN are written as bounds on $\Delta N_{\eff}$, and these can be translated into bounds on the temperature

\begin{equation}
    n_{D} T_{D,BBN}^{4} = \frac{7}{4} \Delta N_{\eff} T_{\gamma,BBN}^{4}.
    \nonumber
\end{equation}

In our scenario, there is no post-BBN reheating mechanism for the dark sector, while the annihilation of electrons and positrons will heat the photons by a factor $(11/4)^{1/3}$. Therefore, our bound can be written as a ratio of the present-day temperatures

\begin{equation}
    g_{D} \bigg(\frac{T_{D}}{T_{\gamma}} \bigg)^{4} = \frac{7}{4} \bigg(\frac{4}{11} \bigg)^{4/3}\Delta N_{\eff}
    \nonumber
\end{equation}

\section{DM interactions with Magnetars}
\label{app:DM-interactions-with-magnetars}

A previous work explored how the stability of magnetar magnetic fields set limits on millicharged magnetic monopoles \cite{Hook:2017vyc}. The authors of this paper considered the Schwinger production of monopoles, which does not require the dark monopoles to comprise dark matter.

It is worth considering the case where the dark monopoles \textit{are} dark matter. For then the magnetic fields no longer need to produce the monopoles from vacuum, in principal allowing larger masses to be probed.

As a first estimate,  suppose all of the monopoles are ionized (this will lead to the maximal possible energy loss for the magnetar). The rate of monopoles entering into the field of influence for the magnetar is

\begin{equation}
    \Gamma_{\rm in} = \frac{\rho_{M}}{m_{M}} v_{M} L^2,
    \nonumber
\end{equation}
where $L$ is the characteristic length of the magnetic fields (about 20 km). Note that because of the difference in signs between coupling to the dark and visible photons, the effective coupling vanishes when $m_{A^{\prime}} L \ll 1$.

Each monopole entering into this region experiences an energy loss of $E_{\rm loss} \simeq B g_{D} \varepsilon L$. The total magnetic energy in the magnetar $E_{\rm tot} \simeq B^2 L^{3}$, with $B \simeq 10^{15} \mathrm{G}$ and these fields are coherent on a time scale  $\tau_{\rm mag} \simeq 10^{4}$ years. As such, this implies
\begin{eqnarray}
   E_{\rm tot} \simeq B^2 L^3 & \gtrsim &\Gamma_{\rm in} E_{\rm loss} \tau_{\rm mag}
   \nonumber \\
  &\simeq & \frac{\rho_{M}}{m_{M}} g_{D} \varepsilon v_{M} B L^3 \tau_{\rm mag},
    \nonumber
\end{eqnarray}
which leads to the following limit 
\begin{equation}
    \varepsilon \lesssim \frac{B m_{M} }{\rho_{M} g_{D} v_{M} \tau_{mag}}.
    \label{eq-eps-sat}
\end{equation}

This constraint is weakened if a significant portion of monopoles remain in bound states as they pass through the region of influence. Despite that, the bound in Eq.~(\ref{eq-eps-sat}) is used as a limiting case and referred to below as $\varepsilon_{sat}$.

\begin{figure}
    \includegraphics[width = 0.45 \textwidth]{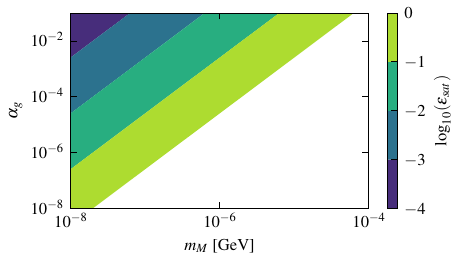}
    \caption{Saturation value of $\varepsilon$ as given by Eq.~(\ref{eq-eps-sat}) as a function of $\alpha_{g}$ and $m_{M}$ for $\rho_{M} = 0.3 \mathrm{GeV \, cm^{-3}}$ and $v_{M} = 10^{-3}$. \label{fig-sat-eps}}
\end{figure}

Bounds on $\varepsilon_{sat}$ are shown in Fig. \ref{fig-sat-eps}, where $m_A \gtrsim L^{-1}$ is assumed. By inspection, constraints only occur for small values of $m_{M}$, and at these masses, magnetars, through the Schwinger mechanism, already strongly constrain this region \cite{Hook:2017vyc}. Furthermore, these values of $m_{M}$ and $\alpha_{g}$ are already excluded through self-interaction bounds (see Figs. \ref{fig-Dark-CMB} and \ref{fig-epsilon-bounds} for examples). The presence of monopole dark matter does not significantly extend the reach possible with magnetars. 

\bibliography{main}

\begin{thebibliography}{44}%
\makeatletter
\providecommand \@ifxundefined [1]{%
 \@ifx{#1\undefined}
}%
\providecommand \@ifnum [1]{%
 \ifnum #1\expandafter \@firstoftwo
 \else \expandafter \@secondoftwo
 \fi
}%
\providecommand \@ifx [1]{%
 \ifx #1\expandafter \@firstoftwo
 \else \expandafter \@secondoftwo
 \fi
}%
\providecommand \natexlab [1]{#1}%
\providecommand \enquote  [1]{``#1''}%
\providecommand \bibnamefont  [1]{#1}%
\providecommand \bibfnamefont [1]{#1}%
\providecommand \citenamefont [1]{#1}%
\providecommand \href@noop [0]{\@secondoftwo}%
\providecommand \href [0]{\begingroup \@sanitize@url \@href}%
\providecommand \@href[1]{\@@startlink{#1}\@@href}%
\providecommand \@@href[1]{\endgroup#1\@@endlink}%
\providecommand \@sanitize@url [0]{\catcode `\\12\catcode `\$12\catcode `\&12\catcode `\#12\catcode `\^12\catcode `\_12\catcode `\%12\relax}%
\providecommand \@@startlink[1]{}%
\providecommand \@@endlink[0]{}%
\providecommand \url  [0]{\begingroup\@sanitize@url \@url }%
\providecommand \@url [1]{\endgroup\@href {#1}{\urlprefix }}%
\providecommand \urlprefix  [0]{URL }%
\providecommand \Eprint [0]{\href }%
\providecommand \doibase [0]{http://dx.doi.org/}%
\providecommand \selectlanguage [0]{\@gobble}%
\providecommand \bibinfo  [0]{\@secondoftwo}%
\providecommand \bibfield  [0]{\@secondoftwo}%
\providecommand \translation [1]{[#1]}%
\providecommand \BibitemOpen [0]{}%
\providecommand \bibitemStop [0]{}%
\providecommand \bibitemNoStop [0]{.\EOS\space}%
\providecommand \EOS [0]{\spacefactor3000\relax}%
\providecommand \BibitemShut  [1]{\csname bibitem#1\endcsname}%
\let\auto@bib@innerbib\@empty
\bibitem [{\citenamefont {'t~Hooft}(1974)}]{tHooft:1974kcl}%
  \BibitemOpen
  \bibfield  {author} {\bibinfo {author} {\bibfnamefont {G.}~\bibnamefont {'t~Hooft}},\ }\href {\doibase 10.1016/0550-3213(74)90486-6} {\bibfield  {journal} {\bibinfo  {journal} {Nucl. Phys. B}\ }\textbf {\bibinfo {volume} {79}},\ \bibinfo {pages} {276} (\bibinfo {year} {1974})}\BibitemShut {NoStop}%
\bibitem [{\citenamefont {Polyakov}(1974)}]{Polyakov:1974ek}%
  \BibitemOpen
  \bibfield  {author} {\bibinfo {author} {\bibfnamefont {A.~M.}\ \bibnamefont {Polyakov}},\ }\href@noop {} {\bibfield  {journal} {\bibinfo  {journal} {JETP Lett.}\ }\textbf {\bibinfo {volume} {20}},\ \bibinfo {pages} {194} (\bibinfo {year} {1974})}\BibitemShut {NoStop}%
\bibitem [{\citenamefont {Nielsen}\ and\ \citenamefont {Olesen}(1973{\natexlab{a}})}]{Nielsen:1973cs}%
  \BibitemOpen
  \bibfield  {author} {\bibinfo {author} {\bibfnamefont {H.~B.}\ \bibnamefont {Nielsen}}\ and\ \bibinfo {author} {\bibfnamefont {P.}~\bibnamefont {Olesen}},\ }\href {\doibase 10.1016/0550-3213(73)90350-7} {\bibfield  {journal} {\bibinfo  {journal} {Nucl. Phys. B}\ }\textbf {\bibinfo {volume} {61}},\ \bibinfo {pages} {45} (\bibinfo {year} {1973}{\natexlab{a}})}\BibitemShut {NoStop}%
\bibitem [{\citenamefont {Nielsen}\ and\ \citenamefont {Olesen}(1973{\natexlab{b}})}]{Nielsen:1973qs}%
  \BibitemOpen
  \bibfield  {author} {\bibinfo {author} {\bibfnamefont {H.~B.}\ \bibnamefont {Nielsen}}\ and\ \bibinfo {author} {\bibfnamefont {P.}~\bibnamefont {Olesen}},\ }\href {\doibase 10.1016/0550-3213(73)90107-7} {\bibfield  {journal} {\bibinfo  {journal} {Nucl. Phys. B}\ }\textbf {\bibinfo {volume} {57}},\ \bibinfo {pages} {367} (\bibinfo {year} {1973}{\natexlab{b}})}\BibitemShut {NoStop}%
\bibitem [{\citenamefont {Nambu}(1974)}]{Nambu:1974zg}%
  \BibitemOpen
  \bibfield  {author} {\bibinfo {author} {\bibfnamefont {Y.}~\bibnamefont {Nambu}},\ }\href {\doibase 10.1103/PhysRevD.10.4262} {\bibfield  {journal} {\bibinfo  {journal} {Phys. Rev. D}\ }\textbf {\bibinfo {volume} {10}},\ \bibinfo {pages} {4262} (\bibinfo {year} {1974})}\BibitemShut {NoStop}%
\bibitem [{\citenamefont {Terning}\ and\ \citenamefont {Verhaaren}(2018)}]{Terning:2018lsv}%
  \BibitemOpen
  \bibfield  {author} {\bibinfo {author} {\bibfnamefont {J.}~\bibnamefont {Terning}}\ and\ \bibinfo {author} {\bibfnamefont {C.~B.}\ \bibnamefont {Verhaaren}},\ }\href {\doibase 10.1007/JHEP12(2018)123} {\bibfield  {journal} {\bibinfo  {journal} {JHEP}\ }\textbf {\bibinfo {volume} {12}},\ \bibinfo {pages} {123} (\bibinfo {year} {2018})},\ \Eprint {http://arxiv.org/abs/1808.09459} {arXiv:1808.09459 [hep-th]} \BibitemShut {NoStop}%
\bibitem [{\citenamefont {Terning}\ and\ \citenamefont {Verhaaren}(2019{\natexlab{a}})}]{Terning:2018udc}%
  \BibitemOpen
  \bibfield  {author} {\bibinfo {author} {\bibfnamefont {J.}~\bibnamefont {Terning}}\ and\ \bibinfo {author} {\bibfnamefont {C.~B.}\ \bibnamefont {Verhaaren}},\ }\href {\doibase 10.1007/JHEP03(2019)177} {\bibfield  {journal} {\bibinfo  {journal} {JHEP}\ }\textbf {\bibinfo {volume} {03}},\ \bibinfo {pages} {177} (\bibinfo {year} {2019}{\natexlab{a}})},\ \Eprint {http://arxiv.org/abs/1809.05102} {arXiv:1809.05102 [hep-th]} \BibitemShut {NoStop}%
\bibitem [{\citenamefont {Terning}\ and\ \citenamefont {Verhaaren}(2019{\natexlab{b}})}]{Terning:2019bhg}%
  \BibitemOpen
  \bibfield  {author} {\bibinfo {author} {\bibfnamefont {J.}~\bibnamefont {Terning}}\ and\ \bibinfo {author} {\bibfnamefont {C.~B.}\ \bibnamefont {Verhaaren}},\ }\href {\doibase 10.1007/JHEP12(2019)152} {\bibfield  {journal} {\bibinfo  {journal} {JHEP}\ }\textbf {\bibinfo {volume} {12}},\ \bibinfo {pages} {152} (\bibinfo {year} {2019}{\natexlab{b}})},\ \Eprint {http://arxiv.org/abs/1906.00014} {arXiv:1906.00014 [hep-ph]} \BibitemShut {NoStop}%
\bibitem [{\citenamefont {Terning}\ and\ \citenamefont {Verhaaren}(2020)}]{Terning:2020dzg}%
  \BibitemOpen
  \bibfield  {author} {\bibinfo {author} {\bibfnamefont {J.}~\bibnamefont {Terning}}\ and\ \bibinfo {author} {\bibfnamefont {C.~B.}\ \bibnamefont {Verhaaren}},\ }\href {\doibase 10.1007/JHEP12(2020)153} {\bibfield  {journal} {\bibinfo  {journal} {JHEP}\ }\textbf {\bibinfo {volume} {12}},\ \bibinfo {pages} {153} (\bibinfo {year} {2020})},\ \Eprint {http://arxiv.org/abs/2010.02232} {arXiv:2010.02232 [hep-th]} \BibitemShut {NoStop}%
\bibitem [{\citenamefont {Okun}(1982)}]{Okun:1982xi}%
  \BibitemOpen
  \bibfield  {author} {\bibinfo {author} {\bibfnamefont {L.~B.}\ \bibnamefont {Okun}},\ }\href@noop {} {\bibfield  {journal} {\bibinfo  {journal} {Sov. Phys. JETP}\ }\textbf {\bibinfo {volume} {56}},\ \bibinfo {pages} {502} (\bibinfo {year} {1982})}\BibitemShut {NoStop}%
\bibitem [{\citenamefont {Holdom}(1986)}]{Holdom:1985ag}%
  \BibitemOpen
  \bibfield  {author} {\bibinfo {author} {\bibfnamefont {B.}~\bibnamefont {Holdom}},\ }\href {\doibase 10.1016/0370-2693(86)91377-8} {\bibfield  {journal} {\bibinfo  {journal} {Phys. Lett. B}\ }\textbf {\bibinfo {volume} {166}},\ \bibinfo {pages} {196} (\bibinfo {year} {1986})}\BibitemShut {NoStop}%
\bibitem [{\citenamefont {Weinberg}(1965)}]{Weinberg:1965nx}%
  \BibitemOpen
  \bibfield  {author} {\bibinfo {author} {\bibfnamefont {S.}~\bibnamefont {Weinberg}},\ }\href {\doibase 10.1103/PhysRev.140.B516} {\bibfield  {journal} {\bibinfo  {journal} {Phys. Rev.}\ }\textbf {\bibinfo {volume} {140}},\ \bibinfo {pages} {B516} (\bibinfo {year} {1965})}\BibitemShut {NoStop}%
\bibitem [{\citenamefont {Zwanziger}(1971)}]{Zwanziger:1970hk}%
  \BibitemOpen
  \bibfield  {author} {\bibinfo {author} {\bibfnamefont {D.}~\bibnamefont {Zwanziger}},\ }\href {\doibase 10.1103/PhysRevD.3.880} {\bibfield  {journal} {\bibinfo  {journal} {Phys. Rev. D}\ }\textbf {\bibinfo {volume} {3}},\ \bibinfo {pages} {880} (\bibinfo {year} {1971})}\BibitemShut {NoStop}%
\bibitem [{\citenamefont {Dirac}(1948)}]{Dirac:1948um}%
  \BibitemOpen
  \bibfield  {author} {\bibinfo {author} {\bibfnamefont {P.~A.~M.}\ \bibnamefont {Dirac}},\ }\href {\doibase 10.1103/PhysRev.74.817} {\bibfield  {journal} {\bibinfo  {journal} {Phys. Rev.}\ }\textbf {\bibinfo {volume} {74}},\ \bibinfo {pages} {817} (\bibinfo {year} {1948})}\BibitemShut {NoStop}%
\bibitem [{\citenamefont {Kobayashi}(1974)}]{Kobayashi:1974ph}%
  \BibitemOpen
  \bibfield  {author} {\bibinfo {author} {\bibfnamefont {M.}~\bibnamefont {Kobayashi}},\ }\href {\doibase 10.1143/PTP.51.1636} {\bibfield  {journal} {\bibinfo  {journal} {Prog. Theor. Phys.}\ }\textbf {\bibinfo {volume} {51}},\ \bibinfo {pages} {1636} (\bibinfo {year} {1974})}\BibitemShut {NoStop}%
\bibitem [{\citenamefont {Berlin}\ \emph {et~al.}(2023)\citenamefont {Berlin}, \citenamefont {Dror}, \citenamefont {Gan},\ and\ \citenamefont {Ruderman}}]{Berlin:2022hmt}%
  \BibitemOpen
  \bibfield  {author} {\bibinfo {author} {\bibfnamefont {A.}~\bibnamefont {Berlin}}, \bibinfo {author} {\bibfnamefont {J.~A.}\ \bibnamefont {Dror}}, \bibinfo {author} {\bibfnamefont {X.}~\bibnamefont {Gan}}, \ and\ \bibinfo {author} {\bibfnamefont {J.~T.}\ \bibnamefont {Ruderman}},\ }\href {\doibase 10.1007/JHEP05(2023)046} {\bibfield  {journal} {\bibinfo  {journal} {JHEP}\ }\textbf {\bibinfo {volume} {05}},\ \bibinfo {pages} {046} (\bibinfo {year} {2023})},\ \Eprint {http://arxiv.org/abs/2211.05139} {arXiv:2211.05139 [hep-ph]} \BibitemShut {NoStop}%
\bibitem [{\citenamefont {Hook}\ and\ \citenamefont {Huang}(2017)}]{Hook:2017vyc}%
  \BibitemOpen
  \bibfield  {author} {\bibinfo {author} {\bibfnamefont {A.}~\bibnamefont {Hook}}\ and\ \bibinfo {author} {\bibfnamefont {J.}~\bibnamefont {Huang}},\ }\href {\doibase 10.1103/PhysRevD.96.055010} {\bibfield  {journal} {\bibinfo  {journal} {Phys. Rev. D}\ }\textbf {\bibinfo {volume} {96}},\ \bibinfo {pages} {055010} (\bibinfo {year} {2017})},\ \Eprint {http://arxiv.org/abs/1705.01107} {arXiv:1705.01107 [hep-ph]} \BibitemShut {NoStop}%
\bibitem [{\citenamefont {Graesser}\ \emph {et~al.}(2022)\citenamefont {Graesser}, \citenamefont {Shoemaker},\ and\ \citenamefont {Arellano}}]{Graesser:2021vkr}%
  \BibitemOpen
  \bibfield  {author} {\bibinfo {author} {\bibfnamefont {M.~L.}\ \bibnamefont {Graesser}}, \bibinfo {author} {\bibfnamefont {I.~M.}\ \bibnamefont {Shoemaker}}, \ and\ \bibinfo {author} {\bibfnamefont {N.~T.}\ \bibnamefont {Arellano}},\ }\href {\doibase 10.1007/JHEP03(2022)105} {\bibfield  {journal} {\bibinfo  {journal} {JHEP}\ }\textbf {\bibinfo {volume} {03}},\ \bibinfo {pages} {105} (\bibinfo {year} {2022})},\ \Eprint {http://arxiv.org/abs/2105.05769} {arXiv:2105.05769 [hep-ph]} \BibitemShut {NoStop}%
\bibitem [{\citenamefont {Rocha}\ \emph {et~al.}(2013)\citenamefont {Rocha}, \citenamefont {Peter}, \citenamefont {Bullock}, \citenamefont {Kaplinghat}, \citenamefont {Garrison-Kimmel}, \citenamefont {Onorbe},\ and\ \citenamefont {Moustakas}}]{CosmicSIDM1}%
  \BibitemOpen
  \bibfield  {author} {\bibinfo {author} {\bibfnamefont {M.}~\bibnamefont {Rocha}}, \bibinfo {author} {\bibfnamefont {A.~H.~G.}\ \bibnamefont {Peter}}, \bibinfo {author} {\bibfnamefont {J.~S.}\ \bibnamefont {Bullock}}, \bibinfo {author} {\bibfnamefont {M.}~\bibnamefont {Kaplinghat}}, \bibinfo {author} {\bibfnamefont {S.}~\bibnamefont {Garrison-Kimmel}}, \bibinfo {author} {\bibfnamefont {J.}~\bibnamefont {Onorbe}}, \ and\ \bibinfo {author} {\bibfnamefont {L.~A.}\ \bibnamefont {Moustakas}},\ }\href {\doibase 10.1093/mnras/sts514} {\bibfield  {journal} {\bibinfo  {journal} {Mon. Not. Roy. Astron. Soc.}\ }\textbf {\bibinfo {volume} {430}},\ \bibinfo {pages} {81} (\bibinfo {year} {2013})},\ \Eprint {http://arxiv.org/abs/1208.3025} {arXiv:1208.3025 [astro-ph.CO]} \BibitemShut {NoStop}%
\bibitem [{\citenamefont {Peter}\ \emph {et~al.}(2013)\citenamefont {Peter}, \citenamefont {Rocha}, \citenamefont {Bullock},\ and\ \citenamefont {Kaplinghat}}]{CosmicSIDM2}%
  \BibitemOpen
  \bibfield  {author} {\bibinfo {author} {\bibfnamefont {A.~H.~G.}\ \bibnamefont {Peter}}, \bibinfo {author} {\bibfnamefont {M.}~\bibnamefont {Rocha}}, \bibinfo {author} {\bibfnamefont {J.~S.}\ \bibnamefont {Bullock}}, \ and\ \bibinfo {author} {\bibfnamefont {M.}~\bibnamefont {Kaplinghat}},\ }\href {\doibase 10.1093/mnras/sts535} {\bibfield  {journal} {\bibinfo  {journal} {Mon. Not. Roy. Astron. Soc.}\ }\textbf {\bibinfo {volume} {430}},\ \bibinfo {pages} {105} (\bibinfo {year} {2013})},\ \Eprint {http://arxiv.org/abs/1208.3026} {arXiv:1208.3026 [astro-ph.CO]} \BibitemShut {NoStop}%
\bibitem [{\citenamefont {Lasenby}(2020)}]{Lasenby:2020rlf}%
  \BibitemOpen
  \bibfield  {author} {\bibinfo {author} {\bibfnamefont {R.}~\bibnamefont {Lasenby}},\ }\href {\doibase 10.1088/1475-7516/2020/11/034} {\bibfield  {journal} {\bibinfo  {journal} {JCAP}\ }\textbf {\bibinfo {volume} {11}},\ \bibinfo {pages} {034} (\bibinfo {year} {2020})},\ \Eprint {http://arxiv.org/abs/2007.00667} {arXiv:2007.00667 [hep-ph]} \BibitemShut {NoStop}%
\bibitem [{\citenamefont {Cruz}\ and\ \citenamefont {McQuinn}(2023)}]{Cruz:2022otv}%
  \BibitemOpen
  \bibfield  {author} {\bibinfo {author} {\bibfnamefont {A.}~\bibnamefont {Cruz}}\ and\ \bibinfo {author} {\bibfnamefont {M.}~\bibnamefont {McQuinn}},\ }\href {\doibase 10.1088/1475-7516/2023/04/028} {\bibfield  {journal} {\bibinfo  {journal} {JCAP}\ }\textbf {\bibinfo {volume} {04}},\ \bibinfo {pages} {028} (\bibinfo {year} {2023})},\ \Eprint {http://arxiv.org/abs/2202.12464} {arXiv:2202.12464 [astro-ph.CO]} \BibitemShut {NoStop}%
\bibitem [{\citenamefont {DeRocco}\ and\ \citenamefont {Giffin}(2025)}]{DeRocco:2024ifs}%
  \BibitemOpen
  \bibfield  {author} {\bibinfo {author} {\bibfnamefont {W.}~\bibnamefont {DeRocco}}\ and\ \bibinfo {author} {\bibfnamefont {P.}~\bibnamefont {Giffin}},\ }\href {\doibase 10.1103/PhysRevD.111.095031} {\bibfield  {journal} {\bibinfo  {journal} {Phys. Rev. D}\ }\textbf {\bibinfo {volume} {111}},\ \bibinfo {pages} {095031} (\bibinfo {year} {2025})},\ \Eprint {http://arxiv.org/abs/2411.11958} {arXiv:2411.11958 [hep-ph]} \BibitemShut {NoStop}%
\bibitem [{\citenamefont {Parker}(1970)}]{Parker:1970xv}%
  \BibitemOpen
  \bibfield  {author} {\bibinfo {author} {\bibfnamefont {E.~N.}\ \bibnamefont {Parker}},\ }\href {\doibase 10.1086/150442} {\bibfield  {journal} {\bibinfo  {journal} {Astrophys. J.}\ }\textbf {\bibinfo {volume} {160}},\ \bibinfo {pages} {383} (\bibinfo {year} {1970})}\BibitemShut {NoStop}%
\bibitem [{\citenamefont {Turner}\ \emph {et~al.}(1982)\citenamefont {Turner}, \citenamefont {Parker},\ and\ \citenamefont {Bogdan}}]{Turner:1982ag}%
  \BibitemOpen
  \bibfield  {author} {\bibinfo {author} {\bibfnamefont {M.~S.}\ \bibnamefont {Turner}}, \bibinfo {author} {\bibfnamefont {E.~N.}\ \bibnamefont {Parker}}, \ and\ \bibinfo {author} {\bibfnamefont {T.~J.}\ \bibnamefont {Bogdan}},\ }\href {\doibase 10.1103/PhysRevD.26.1296} {\bibfield  {journal} {\bibinfo  {journal} {Phys. Rev. D}\ }\textbf {\bibinfo {volume} {26}},\ \bibinfo {pages} {1296} (\bibinfo {year} {1982})}\BibitemShut {NoStop}%
\bibitem [{\citenamefont {Arons}\ and\ \citenamefont {Blandford}(1983)}]{PhysRevLett.50.544}%
  \BibitemOpen
  \bibfield  {author} {\bibinfo {author} {\bibfnamefont {J.}~\bibnamefont {Arons}}\ and\ \bibinfo {author} {\bibfnamefont {R.~D.}\ \bibnamefont {Blandford}},\ }\href {\doibase 10.1103/PhysRevLett.50.544} {\bibfield  {journal} {\bibinfo  {journal} {Phys. Rev. Lett.}\ }\textbf {\bibinfo {volume} {50}},\ \bibinfo {pages} {544} (\bibinfo {year} {1983})}\BibitemShut {NoStop}%
\bibitem [{\citenamefont {{Parker}}(1987)}]{1987ApJ...321..349P}%
  \BibitemOpen
  \bibfield  {author} {\bibinfo {author} {\bibfnamefont {E.~N.}\ \bibnamefont {{Parker}}},\ }\href {\doibase 10.1086/165633} {\bibfield  {journal} {\bibinfo  {journal} {\apj}\ }\textbf {\bibinfo {volume} {321}},\ \bibinfo {pages} {349} (\bibinfo {year} {1987})}\BibitemShut {NoStop}%
\bibitem [{\citenamefont {Kaplan}\ \emph {et~al.}(2010)\citenamefont {Kaplan}, \citenamefont {Krnjaic}, \citenamefont {Rehermann},\ and\ \citenamefont {Wells}}]{Kaplan:2009de}%
  \BibitemOpen
  \bibfield  {author} {\bibinfo {author} {\bibfnamefont {D.~E.}\ \bibnamefont {Kaplan}}, \bibinfo {author} {\bibfnamefont {G.~Z.}\ \bibnamefont {Krnjaic}}, \bibinfo {author} {\bibfnamefont {K.~R.}\ \bibnamefont {Rehermann}}, \ and\ \bibinfo {author} {\bibfnamefont {C.~M.}\ \bibnamefont {Wells}},\ }\href {\doibase 10.1088/1475-7516/2010/05/021} {\bibfield  {journal} {\bibinfo  {journal} {JCAP}\ }\textbf {\bibinfo {volume} {05}},\ \bibinfo {pages} {021} (\bibinfo {year} {2010})},\ \Eprint {http://arxiv.org/abs/0909.0753} {arXiv:0909.0753 [hep-ph]} \BibitemShut {NoStop}%
\bibitem [{\citenamefont {Ma}\ and\ \citenamefont {Bertschinger}(1995)}]{Ma:1995ey}%
  \BibitemOpen
  \bibfield  {author} {\bibinfo {author} {\bibfnamefont {C.-P.}\ \bibnamefont {Ma}}\ and\ \bibinfo {author} {\bibfnamefont {E.}~\bibnamefont {Bertschinger}},\ }\href {\doibase 10.1086/176550} {\bibfield  {journal} {\bibinfo  {journal} {Astrophys. J.}\ }\textbf {\bibinfo {volume} {455}},\ \bibinfo {pages} {7} (\bibinfo {year} {1995})},\ \Eprint {http://arxiv.org/abs/astro-ph/9506072} {arXiv:astro-ph/9506072} \BibitemShut {NoStop}%
\bibitem [{\citenamefont {An}\ \emph {et~al.}(2013{\natexlab{a}})\citenamefont {An}, \citenamefont {Pospelov},\ and\ \citenamefont {Pradler}}]{An:2013yfc}%
  \BibitemOpen
  \bibfield  {author} {\bibinfo {author} {\bibfnamefont {H.}~\bibnamefont {An}}, \bibinfo {author} {\bibfnamefont {M.}~\bibnamefont {Pospelov}}, \ and\ \bibinfo {author} {\bibfnamefont {J.}~\bibnamefont {Pradler}},\ }\href {\doibase 10.1016/j.physletb.2013.07.008} {\bibfield  {journal} {\bibinfo  {journal} {Phys. Lett. B}\ }\textbf {\bibinfo {volume} {725}},\ \bibinfo {pages} {190} (\bibinfo {year} {2013}{\natexlab{a}})},\ \Eprint {http://arxiv.org/abs/1302.3884} {arXiv:1302.3884 [hep-ph]} \BibitemShut {NoStop}%
\bibitem [{\citenamefont {An}\ \emph {et~al.}(2013{\natexlab{b}})\citenamefont {An}, \citenamefont {Pospelov},\ and\ \citenamefont {Pradler}}]{An:2013yua}%
  \BibitemOpen
  \bibfield  {author} {\bibinfo {author} {\bibfnamefont {H.}~\bibnamefont {An}}, \bibinfo {author} {\bibfnamefont {M.}~\bibnamefont {Pospelov}}, \ and\ \bibinfo {author} {\bibfnamefont {J.}~\bibnamefont {Pradler}},\ }\href {\doibase 10.1103/PhysRevLett.111.041302} {\bibfield  {journal} {\bibinfo  {journal} {Phys. Rev. Lett.}\ }\textbf {\bibinfo {volume} {111}},\ \bibinfo {pages} {041302} (\bibinfo {year} {2013}{\natexlab{b}})},\ \Eprint {http://arxiv.org/abs/1304.3461} {arXiv:1304.3461 [hep-ph]} \BibitemShut {NoStop}%
\bibitem [{\citenamefont {Sheldon}(1965)}]{sheldon1965cros}%
  \BibitemOpen
  \bibfield  {author} {\bibinfo {author} {\bibfnamefont {J.~W.}\ \bibnamefont {Sheldon}},\ }\href@noop {} {\emph {\bibinfo {title} {Cross section for impact ionization of H/is/atoms by H/is/atoms near threshold}}},\ \bibinfo {type} {Tech. Rep.}\ (\bibinfo  {institution} {Lewis Research Center},\ \bibinfo {year} {1965})\BibitemShut {NoStop}%
\bibitem [{\citenamefont {Kang}\ and\ \citenamefont {Luty}(2009)}]{Kang:2008ea}%
  \BibitemOpen
  \bibfield  {author} {\bibinfo {author} {\bibfnamefont {J.}~\bibnamefont {Kang}}\ and\ \bibinfo {author} {\bibfnamefont {M.~A.}\ \bibnamefont {Luty}},\ }\href {\doibase 10.1088/1126-6708/2009/11/065} {\bibfield  {journal} {\bibinfo  {journal} {JHEP}\ }\textbf {\bibinfo {volume} {11}},\ \bibinfo {pages} {065} (\bibinfo {year} {2009})},\ \Eprint {http://arxiv.org/abs/0805.4642} {arXiv:0805.4642 [hep-ph]} \BibitemShut {NoStop}%
\bibitem [{\citenamefont {Harnik}\ \emph {et~al.}(2011)\citenamefont {Harnik}, \citenamefont {Kribs},\ and\ \citenamefont {Martin}}]{Harnik:2011mv}%
  \BibitemOpen
  \bibfield  {author} {\bibinfo {author} {\bibfnamefont {R.}~\bibnamefont {Harnik}}, \bibinfo {author} {\bibfnamefont {G.~D.}\ \bibnamefont {Kribs}}, \ and\ \bibinfo {author} {\bibfnamefont {A.}~\bibnamefont {Martin}},\ }\href {\doibase 10.1103/PhysRevD.84.035029} {\bibfield  {journal} {\bibinfo  {journal} {Phys. Rev. D}\ }\textbf {\bibinfo {volume} {84}},\ \bibinfo {pages} {035029} (\bibinfo {year} {2011})},\ \Eprint {http://arxiv.org/abs/1106.2569} {arXiv:1106.2569 [hep-ph]} \BibitemShut {NoStop}%
\bibitem [{\citenamefont {Shellard}(1987)}]{shellard1987cosmic}%
  \BibitemOpen
  \bibfield  {author} {\bibinfo {author} {\bibfnamefont {E.}~\bibnamefont {Shellard}},\ }\href@noop {} {\bibfield  {journal} {\bibinfo  {journal} {Nuclear Physics B}\ }\textbf {\bibinfo {volume} {283}},\ \bibinfo {pages} {624} (\bibinfo {year} {1987})}\BibitemShut {NoStop}%
\bibitem [{\citenamefont {Banks}(2008)}]{banks2008modern}%
  \BibitemOpen
  \bibfield  {author} {\bibinfo {author} {\bibfnamefont {T.}~\bibnamefont {Banks}},\ }\href@noop {} {\emph {\bibinfo {title} {Modern quantum field theory: a concise introduction}}}\ (\bibinfo  {publisher} {Cambridge University Press},\ \bibinfo {year} {2008})\BibitemShut {NoStop}%
\bibitem [{\citenamefont {Caputo}\ \emph {et~al.}(2021)\citenamefont {Caputo}, \citenamefont {Millar}, \citenamefont {O'Hare},\ and\ \citenamefont {Vitagliano}}]{Caputo:2021eaa}%
  \BibitemOpen
  \bibfield  {author} {\bibinfo {author} {\bibfnamefont {A.}~\bibnamefont {Caputo}}, \bibinfo {author} {\bibfnamefont {A.~J.}\ \bibnamefont {Millar}}, \bibinfo {author} {\bibfnamefont {C.~A.~J.}\ \bibnamefont {O'Hare}}, \ and\ \bibinfo {author} {\bibfnamefont {E.}~\bibnamefont {Vitagliano}},\ }\href {\doibase 10.1103/PhysRevD.104.095029} {\bibfield  {journal} {\bibinfo  {journal} {Phys. Rev. D}\ }\textbf {\bibinfo {volume} {104}},\ \bibinfo {pages} {095029} (\bibinfo {year} {2021})},\ \Eprint {http://arxiv.org/abs/2105.04565} {arXiv:2105.04565 [hep-ph]} \BibitemShut {NoStop}%
\bibitem [{\citenamefont {O'Hare}(2025)}]{DPGithub}%
  \BibitemOpen
  \bibfield  {author} {\bibinfo {author} {\bibfnamefont {C.~A.~J.}\ \bibnamefont {O'Hare}},\ }\href@noop {} {\enquote {\bibinfo {title} {Axion limits - dark photon limits},}\ } (\bibinfo {year} {2025}),\ \bibinfo {note} {\url{https://github.com/cajohare/AxionLimits/blob/master/docs/dp.md} [Accessed 2026-01-07]}\BibitemShut {NoStop}%
\bibitem [{\citenamefont {Fields}\ \emph {et~al.}(2020)\citenamefont {Fields}, \citenamefont {Olive}, \citenamefont {Yeh},\ and\ \citenamefont {Young}}]{Fields:2019pfx}%
  \BibitemOpen
  \bibfield  {author} {\bibinfo {author} {\bibfnamefont {B.~D.}\ \bibnamefont {Fields}}, \bibinfo {author} {\bibfnamefont {K.~A.}\ \bibnamefont {Olive}}, \bibinfo {author} {\bibfnamefont {T.-H.}\ \bibnamefont {Yeh}}, \ and\ \bibinfo {author} {\bibfnamefont {C.}~\bibnamefont {Young}},\ }\href {\doibase 10.1088/1475-7516/2020/03/010} {\bibfield  {journal} {\bibinfo  {journal} {JCAP}\ }\textbf {\bibinfo {volume} {03}},\ \bibinfo {pages} {010} (\bibinfo {year} {2020})},\ \bibinfo {note} {[Erratum: JCAP 11, E02 (2020)]},\ \Eprint {http://arxiv.org/abs/1912.01132} {arXiv:1912.01132 [astro-ph.CO]} \BibitemShut {NoStop}%
\bibitem [{\citenamefont {Cyr-Racine}\ and\ \citenamefont {Sigurdson}(2013)}]{Cyr-Racine:2012tfp}%
  \BibitemOpen
  \bibfield  {author} {\bibinfo {author} {\bibfnamefont {F.-Y.}\ \bibnamefont {Cyr-Racine}}\ and\ \bibinfo {author} {\bibfnamefont {K.}~\bibnamefont {Sigurdson}},\ }\href {\doibase 10.1103/PhysRevD.87.103515} {\bibfield  {journal} {\bibinfo  {journal} {Phys. Rev. D}\ }\textbf {\bibinfo {volume} {87}},\ \bibinfo {pages} {103515} (\bibinfo {year} {2013})},\ \Eprint {http://arxiv.org/abs/1209.5752} {arXiv:1209.5752 [astro-ph.CO]} \BibitemShut {NoStop}%
\bibitem [{\citenamefont {Giffin}\ \emph {et~al.}(2025)\citenamefont {Giffin}, \citenamefont {Liu}, \citenamefont {Boucsein}, \citenamefont {Cruz}, \citenamefont {Prabhu}, \citenamefont {Profumo},\ and\ \citenamefont {Roberts}}]{Giffin:2025oqe}%
  \BibitemOpen
  \bibfield  {author} {\bibinfo {author} {\bibfnamefont {P.}~\bibnamefont {Giffin}}, \bibinfo {author} {\bibfnamefont {A.}~\bibnamefont {Liu}}, \bibinfo {author} {\bibfnamefont {J.}~\bibnamefont {Boucsein}}, \bibinfo {author} {\bibfnamefont {A.}~\bibnamefont {Cruz}}, \bibinfo {author} {\bibfnamefont {A.}~\bibnamefont {Prabhu}}, \bibinfo {author} {\bibfnamefont {S.}~\bibnamefont {Profumo}}, \ and\ \bibinfo {author} {\bibfnamefont {M.~G.}\ \bibnamefont {Roberts}},\ }\href@noop {} {\  (\bibinfo {year} {2025})},\ \Eprint {http://arxiv.org/abs/2511.15810} {arXiv:2511.15810 [hep-ph]} \BibitemShut {NoStop}%
\bibitem [{\citenamefont {Murayama}\ and\ \citenamefont {Shu}(2010)}]{Murayama:2009nj}%
  \BibitemOpen
  \bibfield  {author} {\bibinfo {author} {\bibfnamefont {H.}~\bibnamefont {Murayama}}\ and\ \bibinfo {author} {\bibfnamefont {J.}~\bibnamefont {Shu}},\ }\href {\doibase 10.1016/j.physletb.2010.02.037} {\bibfield  {journal} {\bibinfo  {journal} {Phys. Lett. B}\ }\textbf {\bibinfo {volume} {686}},\ \bibinfo {pages} {162} (\bibinfo {year} {2010})},\ \Eprint {http://arxiv.org/abs/0905.1720} {arXiv:0905.1720 [hep-ph]} \BibitemShut {NoStop}%
\bibitem [{\citenamefont {Graesser}\ and\ \citenamefont {Osi{\'n}ski}(2020)}]{Graesser:2020hiv}%
  \BibitemOpen
  \bibfield  {author} {\bibinfo {author} {\bibfnamefont {M.~L.}\ \bibnamefont {Graesser}}\ and\ \bibinfo {author} {\bibfnamefont {J.~K.}\ \bibnamefont {Osi{\'n}ski}},\ }\href {\doibase 10.1007/JHEP11(2020)133} {\bibfield  {journal} {\bibinfo  {journal} {JHEP}\ }\textbf {\bibinfo {volume} {11}},\ \bibinfo {pages} {133} (\bibinfo {year} {2020})},\ \Eprint {http://arxiv.org/abs/2007.07917} {arXiv:2007.07917 [hep-ph]} \BibitemShut {NoStop}%
\bibitem [{\citenamefont {Cielo}\ \emph {et~al.}(2023)\citenamefont {Cielo}, \citenamefont {Escudero}, \citenamefont {Mangano},\ and\ \citenamefont {Pisanti}}]{Cielo:2023bqp}%
  \BibitemOpen
  \bibfield  {author} {\bibinfo {author} {\bibfnamefont {M.}~\bibnamefont {Cielo}}, \bibinfo {author} {\bibfnamefont {M.}~\bibnamefont {Escudero}}, \bibinfo {author} {\bibfnamefont {G.}~\bibnamefont {Mangano}}, \ and\ \bibinfo {author} {\bibfnamefont {O.}~\bibnamefont {Pisanti}},\ }\href {\doibase 10.1103/PhysRevD.108.L121301} {\bibfield  {journal} {\bibinfo  {journal} {Phys. Rev. D}\ }\textbf {\bibinfo {volume} {108}},\ \bibinfo {pages} {L121301} (\bibinfo {year} {2023})},\ \Eprint {http://arxiv.org/abs/2306.05460} {arXiv:2306.05460 [hep-ph]} \BibitemShut {NoStop}%
\end{thebibliography}%
\end{document}